\documentclass[aps, reprint,showpacs,showkeys,pra,longbibliography]{revtex4-1}
\usepackage{dcolumn}
\usepackage[utf8]{inputenc}
\usepackage{amsmath}
\usepackage{amsfonts}
\usepackage{amssymb}
\usepackage{graphicx}
\usepackage{hyperref}
\usepackage{xcolor}
\newcommand{\ket}[1]{\left\vert#1\right\rangle}
\newcommand{\bra}[1]{\left\langle#1\right\vert}

\newcommand{\ketbra}[2]{\left\vert #1 \,\rangle \! \langle #2 \right\vert}

\begin{document}

\title{Describing two-mode squeezed-light experiments without two-mode entanglement or squeezing}

\author{Tam\'iris R. Calixto}
\affiliation{Departamento de F\'isica, Universidade Federal de Minas Gerais, Caixa Postal 701, 30161-970, Belo Horizonte, MG, Brazil}
\author{Pablo L. Saldanha}\email{saldanha@fisica.ufmg.br}
\affiliation{Departamento de F\'isica, Universidade Federal de Minas Gerais, Caixa Postal 701, 30161-970, Belo Horizonte, MG, Brazil}

\date{\today}

\begin{abstract}
In a recent work [Phys. Rev. A \textbf{102}, 053723 (2020)] we have shown that experiments that produce and characterize single-mode light squeezing can be explained in a way where no single-mode squeezed light state is produced in the setup. Here we apply the same ideas to demonstrate that experiments that produce and characterize two-mode light squeezing can also be explained without the production of two-mode squeezed light states. In particular, we show that there is no entanglement between the signal and idler ``twin beam'' modes. This fact may be surprising, since this setup is frequently used to implement entangled-based quantum information protocols such as quantum teleportation. Our work brings an alternative view of the phenomenon. We generalize the Luis and Sánchez-Soto's two-mode relative phase distribution [Phys. Rev. A \textbf{53}, 495 (1996)] to treat four modes, showing that a general physical explanation for the noise reduction in the experiments is a better definition of a phase relation among the four involved optical modes: Signal, idler, and two local oscillators.
\end{abstract}


\maketitle

\section{Introduction}

\par Two-mode squeezed vacuum states, also known as twin beam states, are quantum states of light that may present strong quantum correlations among the modes quadratures \cite{livrogerry,Lvovsky,ou92}. As this class of states have this relevant quantum resource, \textit{i. e.} entanglement, they play a key role in some  quantum information tasks and protocols \cite{braunstein05}, for instance in quantum teleportation \cite{teleportation,PhysRevA.60.937,PhysRevA.62.062307,YAN202143}, quantum dense coding \cite{Ban_1999}, quantum error correction \cite{Walshe2020}, quantum criptography \cite{Eberle:13}, and quantum computing \cite{Zhao2020,Fukui_2022}. The system can also present three-mode entanglement, by considering the twin beam and the laser source \cite{PhysRevLett.97.140504,10.1126/science.1178683}, or high-dimensional quantum entanglement, which permits the implementation of high-dimension quantum information protocols \cite{Chen2014,Kues2019,Erhard2020}. Furthermore, due to their noise reduction properties, the two-mode squeezed vacuum states are specially important within the scope of quantum metrology \cite{PhysRevLett.104.103602,PhysRevA.92.053821,6999929,NatPhotYap}. 

Usually, in experiments of generation and characterization of two-mode squeezed vacuum states, a laser field is considered to be in a coherent state, such that its interaction with a nonlinear medium produces this quantum state of light by non-degenerate parametric down-conversion \cite{livrogerry,Lvovsky}.  A double homodyne detection is then performed in the generated signal and idler fields to characterize the system squeezing \cite{livrogerry,Lvovsky}. However, to consider the state of a laser field as a coherent state is in fact an approximation, since it does not have a known absolute phase. The state of a laser field is better described by a statistical mixture of coherent states \cite{livroqo}, or, similarly, by a statistical mixture of Fock states, not having an optical coherence \cite{molmer}.  This fact generated a debate about if the experimental implementation of quantum teleportation with squeezed light states \cite{teleportation} could really be considered a quantum teleportation protocol \cite{PhysRevLett.87.077903,PhysRevLett.88.027902,PhysRevA.68.042329,twoviews,RevModPhys.79.555}. The eventual conclusion was that in any quantum information protocol with continuous variables a phase reference must be established \cite{twoviews,RevModPhys.79.555}, such that the absence of an absolute phase in a laser field is not a problem and a quantum teleportation protocol was indeed performed in Ref. \cite{teleportation}.

But one should note that to consider that a laser field is in a coherent state is an approximation as good as to consider that it is in a Fock state. The interpretation of different experiments completely change by considering the laser field in a Fock state instead of a coherent state \cite{molmer,saldanha14,calixto20}. These different interpretations for the same experiments certainly brings intriguing questions about the fundamental laws behind the experimental results, enriching the systems understanding. Here we describe two-mode squeezed vacuum experiments using Fock states and statistical mixture of Fock states to represent the laser field. We predict the usual experimental results, but with a completely different interpretation based on interference between the signal and idler fields with the local oscillators fields, similarly to a previous work dealing with single-mode squeezing \cite{calixto20}. In this analysis, we conclude that no two-mode squeezed vacuum state is generated in the experiments and that there is no entanglement between the signal and idler modes. We also generalize the relative phase distribution between two optical modes introduced by Luis and and S\'anchez-Soto \cite{luis96} to define a relative phase distribution among four modes. We then present a general physical explanation for the noise reduction in two-mode squeezing experiments based on this four-mode relative phase distribution involving the signal, idler, and local oscillator fields.

\section{Usual scheme for two-mode squeezing generation and characterization}

\begin{figure}
  \centering
    \includegraphics[width=8.663cm]{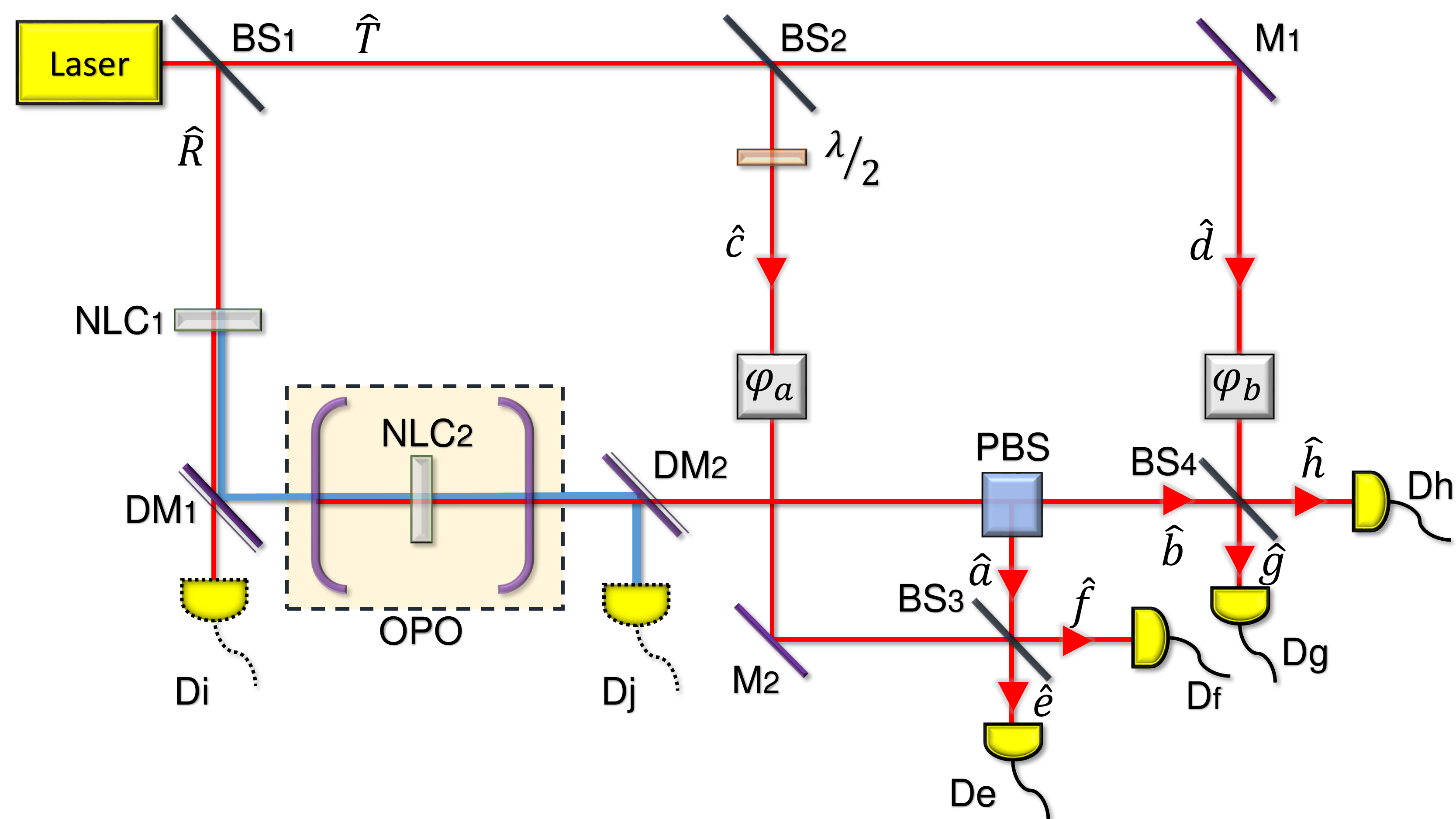}
  \caption{Scheme for producing and characterizing two-mode light squeezing: BS$_{1,2,3,4}$ - beam splitters, NLC$_{1,2}$ - nonlinear crystals, DM$_{1,2}$ - dichroic mirrors, OPO - optical parametric oscillator, $\lambda/2$ - half-wave plate, PBS - polarizing beam splitter, M$_{1,2}$ - mirrors, D$_{\text{e,f,g,h,i,j}}$ - light detectors.}\label{fig:interferometer2mode}
\end{figure}

A scheme for the generation and characterization of two-mode light squeezing is shown in Fig. \ref{fig:interferometer2mode}. Using the traditional treatment, that considers the state of the laser field as a coherent state, two coherent states with smaller amplitudes are transmitted and reflected by the beam splitter BS$_1$. The reflected field, with frequency $\omega$, interacts with the nonlinear crystal NLC$_1$ generating a field with frequency $2\omega$ by second harmonic generation. The dichroic mirror DM$_1$ reflects only the $2\omega$ field to an optical parametric oscillator OPO. The two-mode squeezed vacuum state is generated through the interaction of this field with frequency $2\omega$ with the nonlinear crystal NLC$_2$ inside the OPO, by non-degenerate parametric down-conversion \cite{Lvovsky}. Here we consider that the signal and idler modes have orthogonal polarization, such that the PBS at the exit of the OPO separates them, after the dichroic mirror DM$_2$ to remove the $2\omega$ field. The fields in $a$ (signal) and $b$ (idler) modes are subsequently combined in beam splitters BS$_3$ and BS$_4$ with the local oscillators fields that come from modes $c$ and $d$, assumed to be in intense coherent states $|\beta e^{i\varphi_a}\rangle$ and $|\beta e^{i\varphi_b}\rangle$ respectively (with real $\beta$, $\varphi_a$, and $\varphi_b$). BS$_3$ and BS$_4$ are symmetric 50:50 beam splitters and $\lambda/2$ is a wave plate which lets the polarization of modes $a$ and $c$ identical. Light intensity measurements by detectors D$_\text{e}$, D$_\text{f}$, D$_\text{g}$, and D$_\text{h}$ complete the homodyne detection, these intensities being related to the quadratures of the quantum electromagnetic field in modes $a$ and $b$ \cite{livrogerry,Lvovsky}.
 
A two-mode squeezed vacuum state is produced in modes $a$ and $b$ in the process described in the previous paragraph, which can be written in the Fock basis as  
\begin{align}\label{TMSVS}
    \begin{split}
        &\left|\xi\right>_2=\sum_mC_{m}\left|m\right>_a\left|m\right>_b, \text{ with}\\
        &C_{m}=\operatorname{sech}r(-e^{i\phi}\operatorname{tanh}r)^m,    
    \end{split}
\end{align}
where $r$ and $\phi$ are squeeze parameters that depend on the properties of the OPO and of the incident pump beam \cite{livrogerry,Lvovsky}. The relevant quadratures in this case, which are combinations of the quadratures in modes $a$ and $b$, can be written as $\hat{X}(\theta_a,\theta_b)=\left[\hat{X}(\theta_a)+\hat{X}(\theta_b)\right]/\sqrt{2}$, where $\hat{X}(\theta_j)=(\hat{j}e^{-i\theta_j}+\hat{j}^\dagger e^{i\theta_j})$, with $j=\{a,b\}$. The measured quadratures in the process of homodyne detection can be changed by changing the local oscillator phases $\varphi_a$ and $\varphi_b$ in Fig. 1, since we have $\theta_j=\varphi_j+\pi/2$, again with $j=\{a,b\}$  (the beam splitters BS$_3$ and BS$_4$ include a $\pi/2$ phase in the reflected fields). Considering the state of Eq. (\ref{TMSVS}), it can be shown that the variance of the sum of the photon number differences $\hat{n}_{ef}\equiv\hat{n}_{e}-\hat{n}_{f}$ and $\hat{n}_{gh}\equiv\hat{n}_{h}-\hat{n}_{g}$ is related to the following variance in the quadrature $\hat{X}(\theta_a,\theta_b)$ \cite{livrogerry}:
\begin{equation}\label{eq:8}
        [\Delta(\hat{n}_{ef}+\hat{n}_{gh})]^2\approx8\beta^2[\Delta\hat{X}(\theta_a,\theta_b)]^2.
\end{equation}
By computing $[\Delta\hat{X}(\theta_a,\theta_b)]^2$ for the state of Eq. (\ref{TMSVS}), we obtain \cite{livrogerry}
\begin{align}\label{var_contagens}
\begin{split}
        [\Delta(\hat{n}_{ef}+\hat{n}_{gh})]^2&\approx2\beta^2[\cosh^2{r}+\sinh^2{r}-\\
        &-2\cosh{r}\sinh{r}\cos{(\theta_a+\theta_b-\phi)}].
    \end{split}
\end{align}
Variances for $\hat{n}_{ef}+\hat{n}_{gh}$ smaller than $2\beta^2$ imply variances for $\hat{X}(\theta_a,\theta_b)$ smaller than 1/4, characterizing two-mode squeezing and implying in two-mode entanglement. From Eq. (\ref{var_contagens}), we see that for $r\neq0$ we have two-mode squeezing (when $\phi=\theta_a+\theta_b$) and, consequently, two-mode entanglement. This is the usual interpretation for the phenomenon.

\section{Describing two-mode squeezed-light experiments without two-mode squeezed-light states}

As mentioned before, the laser field state can be written as an incoherent mixture of coherent states with random phases, which is equivalent to an incoherent combination of Fock states \cite{livroqo,molmer}:
\begin{align}\label{laser}
\hat{\rho}_l=\int_0^{2\pi}\frac{d\phi'}{2\pi}\left|\alpha\mathrm{e}^{i\phi'}\right>\left<\alpha\mathrm{e}^{i\phi'}\right|=\sum_{n=0}^{\infty}P_n\left|n\right>\left<n\right|,
\end{align}
with $\alpha$ real and $P_n=\alpha^{2n}e^{-\alpha^2}/(n!)$. But note that the (unknown) laser absolute phase $\phi'$ determines the phase $\phi$ present in the squeezed state of Eq. (\ref{TMSVS}) and also the absolute phases of the local oscillators in modes $c$ and $d$ used in the homodyne detection, such that the intensities measured by the deterctors in Fig. 1 are independent of the laser absolute phase. So, a statistical mixture of coherent state like in Eq. (\ref{laser}) predicts the same experimental results as a pure coherent state $\left|\alpha\mathrm{e}^{i\phi'}\right>$ for the laser field.

The experimental results of the scheme of Fig. 1 should also be predictable with the use of a Fock state to the laser field. As we show in the following, this is indeed the case, but the interpretation of the experimental results is quite different. In this case, the whole setup of Fig. 1 is considered to be a single interferometer with nonlinear elements.

Consider that a Fock state $\left|M\right>$ leaves the laser source in Fig. 1 (we may consider a pulsed laser for convenience). After the BS$_1$ we have an entangled state of the form $\ket{\Psi_1}=\sum_lA_l\ket{l}_R\ket{M-l}_T$, where $A_l$ is a complex coefficient, $l$ is the number of reflected photons and $M-l$ is the number of transmitted photons by the BS$_1$. Second harmonic generation occurs in the NLC$_1$ and photons with frequency $2\omega$ are reflected by the DM$_1$ and set to the OPO, where non-degenerate parametric downconversion occurs. Let us first consider the case where the $2\omega$ pump field incident in the OPO is in a Fock state $\ket{n}_{2\omega}$. In this case, considering a Hamiltonian evolution for the system state in the OPO, where each photon annihilated from the $2\omega$ mode generates one photon in the $a$ mode and one photon in the $b$ mode, we can describe the system state after the OPO as
\begin{equation}\label{OPO2}
	\sum_mC_{n,m}\left|n-m\right>_{2\omega}\left|m\right>_a\left|m\right>_b,
\end{equation}
where $C_{n,m}$ is the probability amplitude of generating $m$ photon pairs in the OPO. If the incident field is in a Fock state $\ket{n'}_{2\omega}$ with $|n'-n|\ll n$, we should have a state with coefficients $C_{n',m}\approx C_{n,m}$, since the the incident pump field would have essentially the same intensity and the probability amplitude of generating $m$ photon pairs should be essentially the same. If, on the other hand, the incident $2\omega$ pump field is in an intense coherent state $\ket{\alpha}_{2\omega}$, this field is not modified by the OPO and the field in modes $a$ and $b$ is given by Eq. (\ref{TMSVS}) \cite{livrogerry,Lvovsky}. 
In this way, we conclude that the coefficients $C_{n,m}$ from Eq. (\ref{OPO2}) can be substituted by the coefficients $C_m$ from Eq. (\ref{TMSVS}), since the probability amplitude for generating $m$ photon pairs in the OPO should be the same if the incident pump field is a Fock state $\ket{n}$ or a coherent state $\ket{\alpha}$ with $|\alpha|^2=n$ large, with a small relative uncertainty in the photon number.

Let us consider that detectors D$_i$ and D$_j$ are positioned as shown in Fig. 1 and count the number of photons $N_i$ and $N_j$ that exit the interferometer by the corresponding modes. The total number of photons left in modes $a$, $b$, $c$, and $d$ is then $N=M-N_i-2N_j$. With these considerations, the system quantum state in modes $a$, $b$, $c$, and $d$ is
\begin{align}\label{eq:9}
    \left|\Psi\right>=\sum_{m, N'}C_{m}Q_{N'}\left|m\right>_a\left|m\right>_b\left|N'-m\right>_{c}\left|N-N'-m\right>_{d},
\end{align}
where the coefficients $C_{m}$ are given by Eq.  (\ref{TMSVS}) and $Q_{N'}$ are complex coefficients. We assume that $m\ll N'$, which is a necessary condition for the process of homodyne detection (the amplitude of the local oscillator fields must be much larger than the one of the signal and idler fields). We also consider a balanced 50:50 beam splitter BS$_2$, such that $N'\approx N-N'\approx N/2$, and  $\sum_{N'}|Q_{N'}|^2=1$.

Using above considerations, in the Appendix it is shown that the variance of the sum of the photon number differences $\hat{n}_{ef}=\hat{n}_{e}-\hat{n}_{f}$ and $\hat{n}_{gh}=\hat{n}_{h}-\hat{n}_{g}$ in the corresponding detectors of Fig. 1 is
\begin{align}\label{eq:10}
    \begin{split}
        [\Delta(\hat{n}_{ef}+\hat{n}_{gh})]^2&\approx N[\cosh^2{r}+\sinh^2{r}-\\
        &-2\cosh{r}\sinh{r}\cos{(\theta_a+\theta_b-\phi)}].
    \end{split}    
\end{align}
This result is equivalent to the predictions of Eq. \eqref{var_contagens} when $N\approx2\beta^2$. 

We thus predict the same experimental results considering the laser field in a Fock state as with the consideration of the laser in a coherent state in the scheme of Fig. 1. But note that from Eq. (\ref{eq:9}) we deduce the following reduced state for modes $a$ and $b$:
\begin{align}\label{eq:11}
    \begin{split}
        \hat{\rho}'_{ab}=\sum_{m=0}^\infty|C_{m}|^2\ketbra{m_a,m_b}{m_a,m_b},
    \end{split}
\end{align}
which is not an entangled or squeezed state, being a statistical mixture of states with equal number of photons in modes $a$ and $b$. So, the experimental results cannot be associated to the generation of two-mode squeezed light states nor to the existence of two-mode entanglement in the system. They are associated to the nonclassical properties of the four-mode state of Eq. (\ref{eq:9}). For each pair of photons that are detected by [D$_e$ or D$_f$] and [D$_g$ or D$_h$], there is a fundamental indistinguishability about if these photons were both generated in the OPO and arrived in modes $a$ and $b$ or if they came from the local oscillator modes and arrived in modes $c$ and $d$. The whole setup of Fig. 1 must be seen as a complex interferometer to describe the experimental results in this case.

If we consider the more realistic state of Eq. (\ref{laser}) for the laser field, with an incoherent combination of Fock states, and also consider that the number of photons in D$_i$ and D$_j$ in Fig. 1 are not measured, the state of Eq. (\ref{eq:9}) will be replaced by a statistical mixture where the total number of photons $N$ is unknown. However, for a large average photon number $\bar{N}$, the relative uncertainty $\Delta N/\bar{N}$ should be small and the experimental prediction for the statistics of Eq. (\ref{eq:10}) should be the same with $\bar{N}$ replacing $N$.

\section{Probability distribution for a phase relation among the four modes}

We have seen that the physical explanation of the experimental results for the generation and characterization of two-mode squeezing in the scheme of Fig. 1 are quite different if we use the coherent states basis to describe the laser field or the Fock states basis to describe this field. A similar situation occurs in the experimental setup that produces and characterizes single-mode squeezing, as we have previously discussed \cite{calixto20}. In this previous work, we've used the two-mode relative phase distribution introduced by Luis and Sánchez-Soto \cite{luis96} to present a general justification for the system noise reduction, in terms of a better definition of the relative phase between the signal and local oscillator fields. We'll do something similar for the two-mode squeezing experiments we discuss here, but first we need to adapt the relative phase distribution definition for the case of four fields: signal, idler, and two local oscillators. It is worth mentioning that the Fisher information obtained from the Luis and Sánchez-Soto's two-mode relative phase distribution was recently shown to be associated to the quantum Fisher information for many pure states useful in quantum metrology \cite{braz22}, showing the value of this phase distribution in the field of quantum metrology.

The starting point is the single-mode phase states of Susskind and Glogower \cite{PhysicsPhysiqueFizika.1.49,livrogerry}:  $\ket{\vartheta}=\frac{1}{\sqrt{2\pi}}\sum_{n=0}^{\infty}e^{in\vartheta}\ket{n}$. These states are considered to have a well defined phase $\vartheta$, although there is no phase operator in quantum optics \cite{mandel}. We then consider phase states for all four modes, fixing the phase of mode $a$ as $\vartheta_a=\vartheta_b+\vartheta_c-\vartheta_d-\Phi$:
\begin{equation}
\begin{split}
    \ket{\Phi^{(4)}}&=\ket{\vartheta_b+\vartheta_c-\vartheta_d-\Phi}_a\ket{\vartheta_b}_b\ket{\vartheta_c}_c\ket{\vartheta_d}_d\\
    &=\frac{1}{(2\pi)^2}\sum_{n_a=0}^{\infty}\sum_{n_b=0}^{\infty}\sum_{n_c=0}^{\infty}\sum_{n_d=0}^{\infty}e^{in_a(\vartheta_d+\vartheta_c-\vartheta_b-\Phi)}\times\\ &\times e^{in_b\vartheta_b}e^{in_c\vartheta_c}e^{in_d\vartheta_d}\ket{n_a}_{a}\ket{n_b}_{b}\ket{n_c}_c\ket{n_d}_d.
\end{split}
\end{equation}
$\ket{\Phi^{(4)}}$ is a state with the following well defined phase relation among the four modes: $(\vartheta_b-\vartheta_d)-(\vartheta_a-\vartheta_c)=\Phi$. Rearranging the exponential terms and performing suitable changes of variables ($n=n_d+n_a$, $l=n_c+n_a$, and $k=n_b-n_a$), this state is written as 
\begin{equation}
    \begin{split}\label{estado de fase}
        \ket{\Phi^{(4)}}&=\frac{1}{(2\pi)^2}\sum_{n_a=0}^\infty\sum_{n=0}^\infty\sum_{l=0}^\infty\sum_{k=-\infty}^\infty e^{-in_a\Phi}e^{ik\vartheta_b}e^{il\vartheta_c}\times\\&\times e^{in\vartheta_d}\ket{n_a}_{a}\ket{k+n_a}_{b}\ket{l-n_a}_c\ket{n-n_a}_d.
    \end{split}    
\end{equation}

 Using the four-mode state of Eq. \eqref{estado de fase}, it is possible to calculate the probability distribution for the four-mode phase relation $\Phi=(\vartheta_b-\vartheta_d)-(\vartheta_a-\vartheta_c)$ inside the interferometer of Fig. \ref{fig:interferometer2mode},
\begin{equation}\label{P-phi}
    P(\Phi)=\int d\vartheta_b\int d\vartheta_c \int d\vartheta_d \bra{\Phi^{(4)}}\hat{\rho}\ket{\Phi^{(4)}},
\end{equation}
where $\hat{\rho}=\ketbra{\Psi}{\Psi}$ is the density operator of the state of Eq. \eqref{eq:9} produced inside the interferometer. Eq. (\ref{P-phi}) corresponds to a generalization of the two-mode relative phase distribution introduced by Luis and Sánchez-Soto \cite{luis96} for the case of four modes. For pure states, the phase probability distribution is given by the expression
\begin{equation}\label{distPfase1}
    P(\Phi)=\int d\vartheta_b\int d\vartheta_c \int d\vartheta_d \left| \langle{\Phi^{(4)}}|{\Psi}\rangle\right|^2.
\end{equation}
Using Eqs. (\ref{estado de fase}) and (\ref{eq:9}), we have
\begin{equation}\label{coef}
\begin{split}
    \langle{\Phi^{(4)}}|{\Psi}\rangle=\frac{1}{(2\pi)^2}\sum_{N',m}e^{-iN'\vartheta_c}e^{-i(N-N')\vartheta_d}e^{im\Phi}Q_{N_{c}}C_m.
\end{split}
\end{equation}
Substituting Eq. \eqref{coef} in Eq. \eqref{distPfase1}, using the relation $\int_0^{2\pi}e^{i(n-n')\phi}d\phi=2\pi\delta_{n,n'}$, and, as mentioned before, assuming that $\sum_{N'}\left|Q_{N'}\right|^2=1$, we finally have the expression for the probability distribution of the four-mode phase relation $\Phi=(\vartheta_b-\vartheta_d)-(\vartheta_a-\vartheta_c)$ for the state of Eq. (\ref{eq:9}): 
\begin{equation}\label{prob.dist.}
\begin{split}
    P(\Phi)=\frac{1}{2\pi}\left|\sum_{m=0}^{\infty}e^{-im\Phi}C_m\right|^2.
\end{split}
\end{equation}

In Fig. \ref{fig:distProb 0.5, 1, 1.5} we plot the phase relation probability distribution of Eq. \eqref{prob.dist.} when the coefficients $C_m$ are given by Eq. \eqref{TMSVS} with  $r=0.5$, $r=1$, and $r=1.5$, always with $\phi=0$. For the three curves, there is one peak in $\Phi=\pi$ and the width of these distributions decrease with the increase of squeeze parameter $r$.
\begin{figure}
  \centering
    \includegraphics[width=8.663cm]{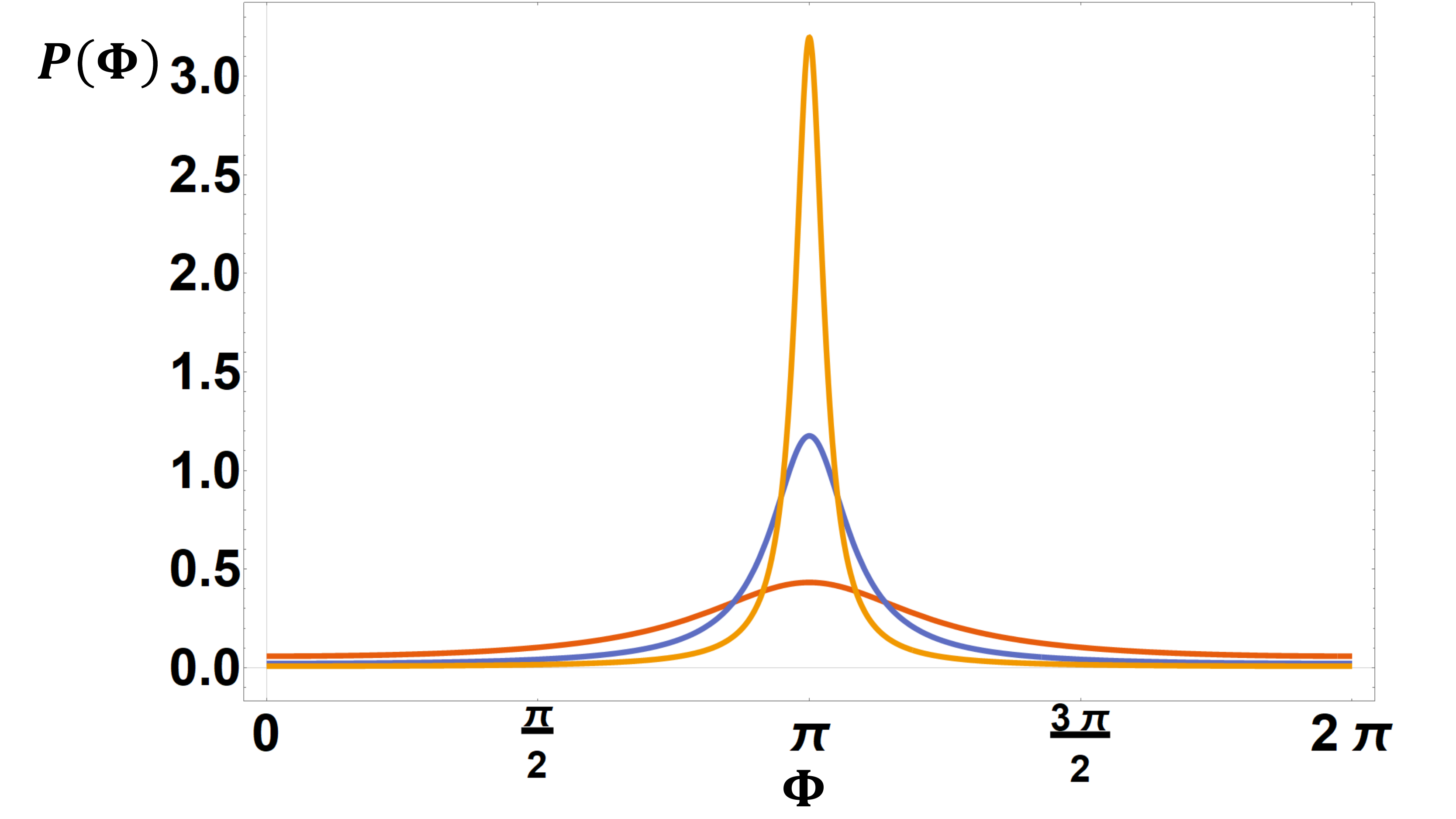}
  \caption{Phase probability distributions $P(\Phi)$ of Eq. (\ref{prob.dist.}) with $C_m$ given by Eq. \eqref{TMSVS} for the squeeze parameters $r=0.5$ (red), $r=1$ (purple) and $r=1.5$ (orange). For all three distributions it was considered $\phi=0$ in Eq. (\ref{TMSVS}) and a superior limit $m_{max}=10000$ for $m$ in the sum of Eq. \eqref{prob.dist.}.}\label{fig:distProb 0.5, 1, 1.5}
\end{figure}

\begin{figure}
  \centering
    \includegraphics[width=8.663cm]{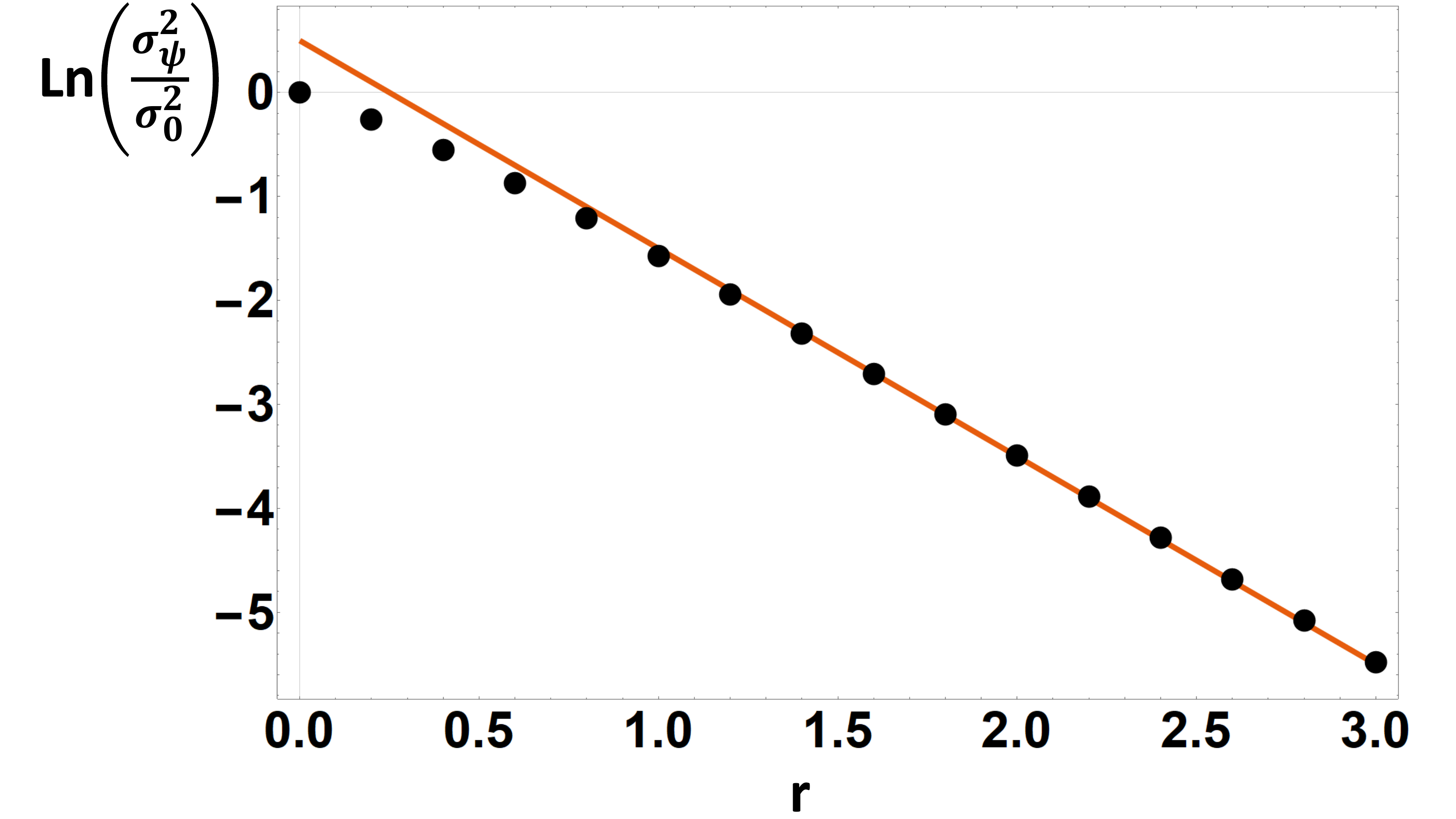}
  \caption{Natural logarithm of the ratio between the variance of $P(\Phi)$ of Eq. (\ref{prob.dist.}) for two-mode squeezing experiments, $\sigma_\psi^2$, and the variance when the modes $a$ and $b$ are in the state $\ket{0,0}$, $\sigma_0^2$, in function of the squeeze parameter $r$. All points were obtained using a superior limit $m_{max}=10000$ for $m$ in Eq. \eqref{prob.dist.} and $\phi=0$ in Eq. (\ref{TMSVS}), considering the phase $\Phi$ between $0$ and $2\pi$. The continuous line corresponds to a dependence of the type $\sigma_\Psi^2=\sigma_0^2e^{-2r+0.5}$.}\label{fig:logvar}
\end{figure}

The variances of these phase probability distributions exhibit an exponentially decreasing behavior with the parameter $r$ for $r>1$. For a better visualization, in Fig. \ref{fig:logvar} is shown the natural logarithm of the ratio between the variance of $P(\Phi)$ for two-mode squeezing experiments, $\sigma_\psi^2$, and the variance when the modes $a$ and $b$ of Fig. \ref{fig:interferometer2mode} are in the two-mode vacuum state $\ket{0,0}$, $\sigma_0^2$, in function of the squeeze parameter $r$. The continuous line shows the function $-2r+0.5$, that fits well the computed points for $r>1$. This behavior is basis independent, which means that if a coherent state is considered to describe the initial state of the laser field, the results for the probability distributions of the considered phase relation among the four modes are essentially the same. The variance of the squeezed quadrature, when we consider a two-mode squeezed vacuum state of Eq. \eqref{TMSVS}, has an analogous behavior, since it decays as $e^{-2r}$ \cite{livrogerry}. 

It can be seen that for small $r$ the variance of the phase probability distributions in Fig. \ref{fig:logvar} are not well fitted by the continuous line. But if we look at Fig. \ref{fig:distProb 0.5, 1, 1.5} and note that the probability distributions for $r<1$ do not decay to zero at $\Phi=0$ and $\Phi=2\pi$, we conclude that the variance is not a good measure of the phase spread in these cases, since $P(\Phi)$ is a $2\pi$-periodic function. Note that the phase $\Phi$ is defined as the sum/difference between 4 phases, each one being in the interval between $-\pi$ and $\pi$, such that it can have values between $-4\pi$ and $+4\pi$. But phase differences $\Phi$ that differ by an integer multiple of $2\pi$ are equivalent. For this reason, if the phase distribution $P(\Phi)$ does not tend to zero in the extremes 0 and $2\pi$ (or $\theta$ and $\theta+2\pi$), its variance is not a good measure of the phase spread. So, it is not a surprise that we do not have a good agreement between the points and the continuous line in Fig. \ref{fig:logvar}  for $r<1$. Unfortunately, we could not find a good measure for the phase spread to be used in this work instead of the variance of $P(\Phi)$. A possible solution would be if we could construct a  phase difference distribution focused on the relative phase itself, without any previous assumption about the absolute phases, as we did in this work having the phase states of Susskind and Glogower \cite{PhysicsPhysiqueFizika.1.49,livrogerry} as a starting point. But unfortunately we were not able to obtain a phase difference distribution on this way. However, note that Fig. \ref{fig:logvar} shows, at least qualitatively, that the variance of $P(\Phi)$ decreases with the increase of the system squeezing for $r<1$. There is, on the other hand, a good agreement between the points and the continuous line in Fig. \ref{fig:logvar} for $r>1$, since the probability distributions are more localized in this regime, such that a quantitative relationship between the variance of $P(\Phi)$ and the system squeezing level can be made in this case. 

We may then conclude that a general explanation for the noise reduction in the experiments depicted in Fig. 1 is a better localization of the relative phase relation $\Phi=(\vartheta_b-\vartheta_d)-(\vartheta_a-\vartheta_c)$ between the four considered modes: signal, idler, and the two local oscillators. This localized phase relation indicates that the fluctuation in $(\vartheta_b-\vartheta_d)$ accompanies the fluctuation in $(\vartheta_a-\vartheta_c)$, generating a high degree of correlation between the photon differences $\hat{n}_{ef}=\hat{n}_{e}-\hat{n}_{f}$ and $\hat{n}_{gh}=\hat{n}_{h}-\hat{n}_{g}$ for large $r$ and suitable phases in the interferometer of Fig. 1. This may be considered the fundamental reason for the noise reduction expressed in Eqs. (\ref{var_contagens}) and (\ref{eq:10}).

\section{Experiments with more than one laser pulse}

In this section we briefly discuss the case where there is a delay of the laser light used as local oscillators in relation to the laser pulse that generates light squeezing. For simplicity, let us consider a (phase-locked) pulsed laser where the first pulse that exits the laser cavity is frequency-doubled and used to pump the OPO, while the second pulse that exits the cavity is used to produce the local oscillators. Assuming that the initial state of the laser cavity is a Fock state with $M$ photons $\ket{M}_\mathrm{las.}$, after the emission of the two pulses the system state is $\sum_{k,l}A_{k,l}\ket{k}_1\ket{l}_2\ket{M-k-l}_\mathrm{las.}$, disregarding any photon generation in the laser cavity in this period in our toy model. $\ket{k}_1$ is a state with $k$ photons in the first pulse and $\ket{l}_2$ a state with $l$ photons in the second pulse. Following the steps that leaded us to Eq. (\ref{eq:9}), the system state in modes $a$, $b$, $c$, and $d$ of  Fig. 1, plus the laser cavity mode, is
\begin{equation}\label{aux}
	\left|\Psi\right>=\sum_{m,l,p}C_{m}B_{l,p}\left|m\right>_a\left|m\right>_b\left|p\right>_{c}\left|l-p\right>_{d}\ket{M'-l-2m}_\mathrm{las.},
\end{equation}
where $M'=M-N_i-2N_j$ is the system total number of photons after the subtraction of the $N_i$ photons detected by D$_i$ and the $N_j$ photons detected by D$_j$ in Fig. 1. Consider that the number of photons in the laser cavity is also measured and the value $M_\mathrm{las.}$ is found. The total number of photons left in the system is then $N=M'-M_\mathrm{las.}$ and we must impose the condition $M'-l-2m=M_\mathrm{las.}$, or $l=N-2m$, in Eq. (\ref{aux}). The system state then becomes
\begin{equation}\label{aux2}
	\left|\Psi\right>=\sum_{m,l,p}C_{m}B_{N-2m,p}\left|m\right>_a\left|m\right>_b\left|p\right>_{c}\left|N-2m-p\right>_{d}.
\end{equation}
The above state is equivalent to the one of Eq. (\ref{eq:9}) if we define $p=N'-m$ and note that $B_{N-2m,p}\approx B_{N,p}$ for $N\gg 2m$, such that the same experimental predictions are made. In particular, the prediction for the variance of the sum of the photon number differences $\hat{n}_{ef}=\hat{n}_{e}-\hat{n}_{f}$ and $\hat{n}_{gh}=\hat{n}_{h}-\hat{n}_{g}$ in the corresponding detectors of Fig. 1 is is the same as in Eq. (\ref{eq:10}). Again, if we consider the more realistic state of Eq. (\ref{laser}) for the laser field, and also that the number of photons in the laser cavity field, in D$_i$, and in D$_j$ are not measured, the state of Eq. (\ref{aux2}) will be replaced by a statistical mixture where the total number of photons $N$ is unknown, but with a small relative variation in the photon number. In this case, the experimental prediction for the statistics of Eq. (\ref{eq:10}) should be the same with the average photon number replacing $N$, as before.

Van Enk and Fucks \cite{PhysRevLett.88.027902} and Pegg \cite{pegg09} have explicitly shown how there can be phase coherence between the light that exits a laser cavity at different times for continuous \cite{PhysRevLett.88.027902} and pulsed \cite{pegg09} lasers. Even if the initial cavity state has a completely undetermined phase, being a statistical mixture of Fock states, there is a well defined phase relation between the different pulses of a phase-locked laser field \cite{pegg09}. The phase of each pulse is not well defined, with the expected value of the electric field operator being zero at all times, but the phase relation between consecutive pulses is well defined. This well defined phase relation is associated to the correlation between the number of photons of the different laser pulses and the laser cavity fields \cite{pegg09}. This correlation is present in the discussion of the previous paragraph, being essential for correctly predicting the experimental results on two-mode squeezing. So, homodyne measurements work even if the local oscillators fields do not come from the same pulse that generated the squeezed light field.

\section{Conclusion}
\par To conclude, we have investigated the physical interpretation of experiments that generate and characterize two-mode squeezed light taking into account the fact that the laser field used in the experiments is not a coherent state, but an incoherent combination of coherent states with random phases, which is equivalent to an incoherent combination of Fock states. Interestingly, the physical explanations of the experimental results are quite different depending on which basis we describe the laser field state. In the Fock basis description, no two-mode squeezed vacuum state is generated and we conclude that there is no entanglement between the signal and idler modes. It is important to mention that, although there is no entanglement when we consider only the signal and idler modes, there may be entanglement when the local oscillators are taken into account, but this four-mode entanglement analysis is outside the scope of the present paper. We have extended the Luis and Sánchez-Soto's  two-mode relative phase distribution \cite{luis96} to describe a phase relation between the four relevant modes in the experiment: Signal, idler, and two local oscillators. We have seen that a general physical explanation for the noise reduction in the system is the reduction in the variance of a phase relation probability distribution among these four modes with the increase of the squeeze parameter $r$. We hope our work brings fundamental clarifications to this important physical setup in quantum optics and quantum information. 

The authors acknowledge Marcelo Terra Cunha for useful discussions. This work was supported by the Brazilian agencies CNPq (Conselho Nacional de Desenvolvimento Cient\'ifico e Tecnol\'ogico), CAPES (Coordenação de Aperfeiçoamento de Pessoal de Nível Superior), and FAPEMIG (Fundação de Amparo à Pesquisa do Estado de Minas Gerais).

\appendix* \section{}

In this Appendix we deduce the variance of $(\hat{n}_{ef}+\hat{n}_{gh})$ from Eq. (\ref{eq:10}). We employ the the usual expression $[\Delta(\hat{n}_{ef}+\hat{n}_{gh})]^2=\left<(\hat{n}_{ef}+\hat{n}_{gh})^2\right>-\left<(\hat{n}_{ef}+\hat{n}_{gh})\right>^2$, with $\hat{n}_{ef}=\hat{n}_{e}-\hat{n}_{f}$ and $\hat{n}_{gh}=\hat{n}_{h}-\hat{n}_{g}$. The annihilation operators are   $\hat{e}=(\hat{a}+i\hat{c}e^{i\varphi_a})/\sqrt{2}$, $\hat{f}=(i\hat{a}+\hat{c}e^{i\varphi_a})/\sqrt{2}$, $\hat{g}=(i\hat{b}+\hat{d}e^{i\varphi_b})/\sqrt{2}$ and  $\hat{h}=(\hat{b}+i\hat{d}e^{i\varphi_b})/\sqrt{2}$.
So we have $(\hat{n}_{ef}+\hat{n}_{gh})=\hat{a}^\dagger\hat{c}e^{i\theta_a}+\hat{c}^\dagger\hat{a}e^{-i\theta_a}+\hat{b}^\dagger\hat{d}e^{i\theta_b}+\hat{d}^\dagger\hat{b}e^{-i\theta_b}$, where $\theta_i=\varphi_i+\pi/2$ for $i=\{a,b\}$. Using the state of Eq. \eqref{eq:9}, making the approximation $N'\approx N-N'\approx N/2$, and considering $\sum_{N'}|Q_{N'}|^2=1$, we obtain
\begin{align}\label{A2-1}
    \begin{split}
        \left<[\Delta(\hat{n}_{ef}+\hat{n}_{gh})]^2\right>&\approx N\sum_{m=0}^\infty\big[\left|C_m\right|^2(2m+1)+ \\
        &+C_m^*C_{m+1}e^{-i(\theta_a+\theta_b)}(m+1)+\\
        &+C_{m+1}^*C_{m}e^{i(\theta_a+\theta_b)}(m+1)\big].
    \end{split}
\end{align}
With the explicit form of the coefficients $C_m$ given by the Eq. \eqref{TMSVS}, the expression above is written as
\begin{align}\label{A2-2}
    \begin{split}
        \left<[\Delta(\hat{n}_{ef}+\hat{n}_{gh})]^2\right>&\approx N\sum_{m=0}^\infty \operatorname{sech}^2r\big[(\operatorname{tanh}r)^{2m}(2m+1)-\\
        &-2(\operatorname{tanh}r)^{2m+1}(m+1)\cos{(\theta_a+\theta_b-\phi)}\big].
    \end{split}
\end{align}

For $x^2<1$, the formula of a geometric series is
\begin{align}\label{A2-3}
    \frac{1}{1-x^2}=\sum_{n=0}^\infty x^{2n},
\end{align}
from which we obtain
\begin{align}\label{A2-4}
    \frac{d}{dx}\left(\frac{1}{1-x^2}\right)=\sum_{n=0}^\infty2nx^{2n-1}.
\end{align}
We can apply Eq. \eqref{A2-3} and Eq. \eqref{A2-4} in Eq. \eqref{A2-2} with $x=\tanh{r}$ to show the result seen in Eq. \eqref{eq:10}, using also the relations $\operatorname{tanh}r={\sinh{r}}/{\cosh{r}}$, $\operatorname{sech}r={1}/{\cosh{r}}$, and $\cosh^2{r}-\sinh^2{r}=1$.


%

\end{document}